\documentclass[twocolumn,showpacs,prb,amsmath,amssymb,longbibliography]{revtex4-2}
\usepackage{graphicx}
\usepackage[utf8]{inputenc}

\begin{document}

\title{Binding energy of polaronic trions and biexcitons in CsPbBr$_3$ nanocrystals}

\author{Jos\'e L. Movilla}
\affiliation{Dept. d'Educaci\'o i Did\`actiques Espec\'ifiques, Universitat Jaume I, 12080, Castell\'o, Spain}

\author{Josep Planelles, Juan I. Climente}
\affiliation{Dept. de Qu\'imica F\'isica i Anal\'itica, Universitat Jaume I, 12080, Castell\'o, Spain}
\email{climente@uji.es}
\date{\today}

\begin{abstract}
The effect of polaron formation on the ground state of excitons, trions and biexcitons confined in CsPbBr$_3$ nanocrystals
is studied in the framework of effective mass Hamiltonians, using a Haken-like (Bajaj) potential for carrier-phonon coupling.
The binding energy of trions agrees well with that observed in experiments, with position-dependent dielectric screening
playing a significant role. 
For biexcitons, however, neither polaronic effects, nor dielectric confinement, nor electronic correlations
--here accounted for with a variational Quantum Monte Carlo method--
 suffice to explain the large binding energies reported by single nanocrystal spectroscopy experiments.
This result reinforces the hypothesis that biexcitons polarize the perovskite lattice differently from
excitons and trions.
\end{abstract}

\maketitle

\section{Introduction}

CsPbX$_3$ (X=Cl,Br,I) nanocrystals (NCs) combine the outstanding optoelectronic properties of lead halide perovskites
with the tunability of quantum and dielectric confinement.\cite{AkkermanNM,AkkermanJPCL,ButkusCM,CaicedoAFM}
Understanding the interactions among particles confined inside such systems is a prerequisite for eventual applications.
In the last years, a large number of experimental studies have gone beyond exciton (X) complexes to address trions 
(X$^*$) and biexcitons (BX).\cite{TamaratNC,FuNL,BeckerNAT,WangAM,MakarovNL,CastanedaACS,AneeshJPCc,YumotoJPCL,AshnerACSener,HuangJPCL,ShenJPCc,DanaACS,PooniaPRB,LubinACS,AmaraNL,ZhuAM,ChoACS,KazesNL}
The interest of X$^*$ is largely related to the fact that they combine the optical response of X
with the high sensitivity to external electric fields of charged species, which is convenient for field-modulated devices.\cite{ZieglerAM}
BX in turn are involved in several potential applications, including lasers, LEDs and quantum light sources.\cite{WangAM,ZhuAM}
Prospects of BX exploitation in CsPbX$_3$ NCs are backed up by near-unity quantum yields,
which result from the fast radiative rates outcompeting Auger processes.\cite{UtzatSCI}

The binding energy of X$^*$ and BX is a key figure quantifying the spectral shift of these species with respect to X,
as well as the strength of the charge-X (for X$^*$) and X-X (for BX) interactions.
Different experiments have provided very different estimates, especially in the case of BX, depending on the methodology, 
temperature and size inhomogeinity of the NCs.\cite{LubinACS} In view of this, the most reliable measurements are arguably
those provided by single NC emission spectroscopy at cryogenic temperatures.
For CsPbBr$_3$ NCs with lateral sizes between 5 and 20 nm, these measurements reveal 
trion binding energies ($\Delta_{X^*}$) in the range of $7-25$ meV,
and biexciton binding energies ($\Delta_{BX}$) in the range of $25-40$ meV.\cite{TamaratNC,FuNL,BeckerNAT,ZhuAM,ChoACS,AmaraNL}

 Attempts to rationalize these values have been made using different theoretical models, 
 with different descriptions of the dielectric screening and carrier-carrier interactions. 
 In all cases, trion binding energies were in fair agreement with experiments, but 
 biexciton ones were substantially and systematically lower.\cite{ZhuAM,ChoACS,NguyenPRB}
 This puzzling result prompted some authors to postulate that, in inorganic lead halide NCs,
 BX feel a lower effective dielectric constant than X.\cite{ZhuAM} The same hypothesis
 was adopted in later studies, for simulations to reproduce spectroscopic data.\cite{ChoACS} 

 In this paper, we aim at providing more solid understanding on the origin of the unusually 
 large $\Delta_{BX}$ values. To that end, we carry out a theoretical description of X, X$^*$
 and BX in cuboidal CsPbBr$_3$ NCs. As in previous works, we rely on effective mass models to 
 account for the influence of quantum confinement\cite{ZhuAM,ChoACS,NguyenPRB}, 
 and on image charges to account for dielectric confinement.\cite{ChoACS}
 Two relevant refinements are however introduced in the model. 
 The first one refers to electronic correlations. 
 Because CsPbBr$_3$ NCs are in the intermediate to weak confinement regime, 
 carrier-carrier correlations are known to play a major role.\cite{BlundellPRB}
 Here we go beyond second-order perturbation\cite{NguyenPRB} and configuration interaction
 methods,\cite{ZhuAM,ChoACS}
 as they have proved insufficient in other weakly confined nanostructures.\cite{PlanellesTCA}
 Rather, we implement a variational Quantum Monte Carlo approximation.\cite{PlanellesCPC,MaciasNS}
  
 The second refinement refers to polaronic effects. 
 Lead halide perovskites are known to have a soft lattice, such that carrier-lattice coupling becomes
 a significant physical factor in shaping the electronic structure.
 In bulk, carriers couple to optical phonons forming polarons,\cite{PuppinPRL,DaiPRL} 
 which lead to a non-hydrogenic exciton spectrum.\cite{MenendezPSS,BaranowskiAEM,BaranowskiACSener}
 Polaron signatures are also expected in NCs, but 
 their effect is not clear yet. Based on different experimental results, some authors have
 suggested polarons may be responsible for large and attractive $\Delta_{BX}$,\cite{DanaACS}
 large and repulsive $\Delta_{BX}$,\cite{PooniaPRB}  and even break the crystal symmetry.\cite{RossiACS}
 On the theoretical side, Park and Limmer used path integral molecular quantum dynamics to show
 that polaronic effects are relevant in determining $\Delta_X$ but less so for $\Delta_{BX}$.\cite{ParkPRM}
 To shed more light on this matter, here we complement the study of polaronic effects by means of 
 simpler and intuitive Haken-like potential,
 which takes into account the different dielectric screening at short and long distances from
 carriers.\cite{MenendezPSS,BaranowskiAEM}
 This approximation has proved successful in reproducing excitonic features
 of bulk lead halide perovskites.\cite{BaranowskiACSener} Similar polaronic potentials have
 proved successful in reproducing the binding energy of X, X$^*$ and BX in layered
 halide perovskites and nanoplatelets.\cite{MovillaNSa,ClimenteJPCL,ClimenteJPCc}
 We shall extend the study to BX and trions in CsPbBr$_3$ NCs and compare our findings 
 with experiments and with Ref.~\onlinecite{ParkPRM}.

 Our calculations show that polaronic effects have a sizable influence on charge-X
 and X-X interactions. $\Delta_{X^*}$ values in close agreement with spectroscopic data
 are obtained, with no need for an effective dielectric constant inside the NC. 
 Nonetheless, $\Delta_{BX}$ remain lower than those observed in experiments.
 Agreement is only reached if one assumes that X and BX polarize the lattice differently,
 for example by imposing greater polaron radius for BX.

\section{Theoretical Model}
\label{s:theo}

We calculate the ground state energy and wave function of excitons, negative trions and biexcitons confined
in cuboidal CsPbBr$_3$ NCs.
The Hamiltonians are based on $k\cdot p$ theory for two uncoupled (conduction and valence) bands:

\begin{eqnarray} 
	\label{eq:HX}
	H_{X} &=&\sum_{i=e,h}  T_i
	+ V_c(\mathbf{r}_{e},\mathbf{r}_{h}), \\
	\label{eq:HX*}
	H_{X^*} &=&\sum_{i=e_1,e_2,h}  T_i
	+ \sum_{i=e_1,e_2} V_c(\mathbf{r}_{h},\mathbf{r}_{i}) 
        +V_c(\mathbf{r}_{e_1},\mathbf{r}_{e_2}), \\ 
	\label{eq:BX}
	H_{BX} &=&\sum_{i=e_1,e_2,h_1,h_2} T_i  
	+ \sum_{i=e_1,e_2}\sum_{j=h_1,h_2}  V_c(\mathbf{r}_{j},\mathbf{r}_{i}) \nonumber \\ 
	&+& V_c(\mathbf{r}_{e_1},\mathbf{r}_{e_2})
	+V_c(\mathbf{r}_{h_1},\mathbf{r}_{h_2}). 
\end{eqnarray}
\noindent Here, $T_i=(\mathbf{p}_i^2/(2m_i)+ V_i )$ is the single-particle (kinetic and potential) energy operator,
with $m_i$ the effective mass and $\mathbf{p}_i$ the momentum operator.
%
The single-particle potential is $V_i=V_i^{conf}+V_i^{self}$, where
$V_i^{conf}$ is the spatial confining potential. In our model, 
we describe the NCs as a cuboid with lateral dimension $L$. 
For holes, we take $V_h^{conf}=0$ inside the cuboid and infinite outside it.
For electrons, $V_e^{conf}=E_{gap}$ (the bulk band gap) 
inside the cuboid and infinite outside it.
$V_i^{self}$ is the self-energy potential.
 $V_c(\mathbf{r}_{i},\mathbf{r}_{j})$ terms represent the Coulomb interaction between carriers.
 Both $V_i^{self}$ and $V_c(\mathbf{r}_i,\mathbf{r}_j)$ account for dielectric confinement
 by using quantum box image charges, with inclusion of long-range\cite{TakagaharaPRB} and short-range\cite{MovillaNSa} interactions.
 They are based on a Bajaj potential\cite{BajajSSC}, a Haken-like model of polaronic interactions.\cite{BaranowskiAEM} 
 %
The validity of these models in CsPbBr$_3$ is supported by recent results showing that 
descriptions of carrier-lattice interaction using harmonic phonons provide comparable accuracy 
to fully anharmonic phonons in determining excitonic energies.\cite{ParkPRM}

 To assist with the interpretation of the results, it is worth introducing here the basic Bajaj potential
 (for bulk systems, prior to inclusion of quantum and dielectric confinement).
 In atomic units, the potential exerted by a source charge $i$ on a test charge $j$ reads:\cite{BajajSSC}
\begin{equation}
\label{eq02}
V_{ij}^{bulk}(r) = \frac{q_i q_j}{\varepsilon_s \, r}
	+ \left( \frac{\varepsilon_\infty}{\varepsilon_s} \right)^\gamma \,
	\frac{q_i q_j}{\varepsilon^* \,r}   \, \frac{e^{-\beta_i\, r}+e^{-\beta_j\, r}}{2}.
\end{equation}
\noindent  Here, $r$ is the distance between charges, $q$ the elementary charge (positive or negative) 
and $\varepsilon_s$ the static dielectric constant.
 The term $\left( \frac{1}{\varepsilon_{\infty}}- \frac{1}{\varepsilon_s}\right) = \frac{1}{\varepsilon^*}$
represents the ionic screening of carriers, with $\varepsilon_\infty$ the high frequency dielectric constant and
 $\beta_i=1/l_i$, with $l_i$ the polaron radius of carrier $i$. 
 The latter is defined as:
 \begin{equation}
	 \label{eq:l}
	 l_i = \sqrt{\frac{\hbar^2}{2 \, m_i^b \, E_{LO}}},
 \end{equation}
 \noindent with $E_{LO}$ the LO phonon energy and $m_i^b$ the bare effective mass,
 which relates to the polaronic one as $m_i = m_i^b \, (1 + \alpha/6)$, $\alpha$ being the Fr\"ohlich coupling constant.
One should note that the first term in Eq.~(\ref{eq02}) is a Coulomb interaction with full dielectric screening (electronic plus ionic, $\varepsilon_s$),
which prevails at long distances, $r \gg l_i$.
On the other hand, the second term is a short-range (Yukawa) one, which starts prevailing at distances $r \lesssim l_i$,
where the ionic contribution to dielectric screening is lost ($\varepsilon_\infty$ dominates).
The term $(\varepsilon_\infty/\varepsilon_s)^\gamma$, with $\gamma$ a material dependent parameter, 
is the distinct trait of Bajaj potential as compared to the original Haken one.
The goal of this term is to improve (phenomenologically) the estimates of exciton binding energies.\cite{BajajSSC}
It should be noted that the use of the bajaj potential implicitly assumes weakly-confined systems. 
This is consistent with the nanocrystals we study, 
where the energy of all phonon modes observed are size-independent (that is, bulk-like).\cite{AmaraNL}

In NCs, the potential Eq.~(\ref{eq02}) is renormalized due to the dielectric confinement.
The modified potential can be expressed as follows:\cite{MovillaNSa}
\begin{equation}
\label{eqVeh}
V_{ij}=\sum_{k=C,Y} \; \sum_{l=-\infty}^{\infty} \sum_{m=-\infty}^{\infty} \sum_{n=-\infty}^{\infty} \frac{q_i q_j\,q_{lmn,k}\,f_{lmn,k}({\mathbf r}_i,{\mathbf r}_j)}{r_{lmn}}.
\end{equation}
\noindent Here,
\begin{equation}
\label{eqrn}
r_{lmn} = \left [ (x_i-x_{jl})^2 + (y_i-y_{jm})^2 + (z_i-z_{jn})^2 \right ]^{1/2},
\end{equation}
\begin{eqnarray}
    &&  (x_{jl},y_{jm},z_{jn}) =  \nonumber \\
    &&  \left [ lL+(-1)^l x_j , mL+(-1)^m y_j , nL+(-1)^n z_j \right ],
\label{eqrn2}
\end{eqnarray}
\noindent and
\begin{equation}
	q_{lmn,k} = \frac{1}{\varepsilon_k}\, \left [ \frac{\varepsilon_k - \varepsilon_2}{\varepsilon_k + \varepsilon_2} \right ]^{\left | l \right | + \left | m \right | + \left | n \right |},
\end{equation}
\noindent  with $\varepsilon_C = \varepsilon_s$ (the static dielectric constant of the perovskite material), 
$\varepsilon_Y = \varepsilon^*$ (ionic correction of the perovskite), and $\varepsilon_2$ the dielectric constant of the outer medium.
The functions $f_{lmn,k}$ are given by:
\begin{eqnarray}
\label{eqf}
	f_{lmn,C}({\mathbf r}_i,{\mathbf r}_j) &=& 1, \\
	f_{lmn,Y}({\mathbf r}_i,{\mathbf r}_j) &=& \frac{1}{2} \left( \frac{\varepsilon_{\infty}}{\varepsilon_s} \right)^{\gamma} \left ( e^{-\beta_i r_{lmn}} + e^{-\beta_j r_{lmn}} \right ).
\end{eqnarray}

 The single-particle self-polarization potentials $V_i^{self}$ can be obtained from Eq.~(\ref{eqVeh}) by setting ${\mathbf r}_i = {\mathbf r}_j$, eliminating the $l=m=n=0$ term (which represents the interaction of a charge carrier with itself), and dividing by 2 as corresponds to a self-energy.

Hamiltonians (\ref{eq:HX})-(\ref{eq:BX}) are integrated using Quantum Monte Carlo.\cite{CeperleyPRB,NeedsJPCM}
Details on the evaluation of local energy and computational aspects can be found in Refs.~\onlinecite{PlanellesCPC,MaciasNS,ClimenteJPCc}. 
%
The variational functions we integrate are built as the product of non-interacting functions 
(describing the confinement-defined limit) 
and appropriate correlation factors (describing the bulk limit). 
It is the interplay between confinement and carrier interactions that determines 
the optimal value of the (isotropic) variational parameters.
For X, the unnormalized function reads:
\begin{equation}
\label{eqtrial}
	\Psi_X= \Phi_e(\mathbf{r}_e) \Phi_h(\mathbf{r}_h) \, e^{-Z r_{eh}}.
\end{equation}

\begin{figure}[h]
\includegraphics[width=6.0cm]{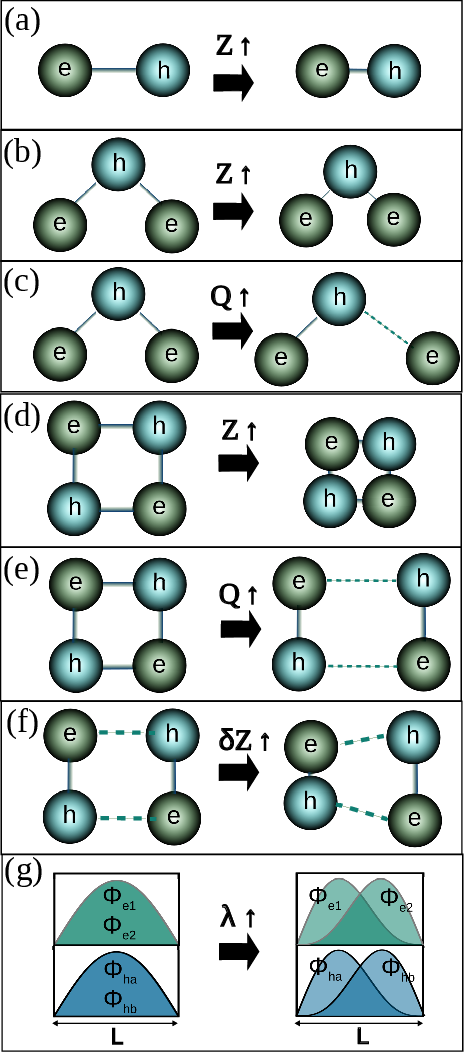}
	\caption{Sketch of the effect of different variational parameters on 
	(a) $\Psi_X$, (b-c) $\Psi_{X^*}$, (d-g) $\Psi_{BX}$.
	For illustration purposes, 2D charge distributions are used.
	$Z$ determines the electron-hole distances.
	$Q$ enables gradual dissociation of X$^*$ and BX.
	For BXs, $\delta Z$ introduces asymmetry between two partially dissociated Xs.
	$\lambda$ introduces asymmetry in the envelope function of
	carriers with the same sign.}
\label{fig1}
\end{figure}

\noindent Here, $\Phi_e(\mathbf{r}_e)$ and $\Phi_h(\mathbf{r}_h)$ are the non-interacting electron and hole (particle-in-box) states,
which account for quantum confinement. The exponential term is a Slater correlation factor, 
which captures excitonic interactions. $Z$, the inverse of the exciton Bohr radius, is the parameter to optimize variationally.
Figure \ref{fig1}(a) depicts the qualitative effect of increasing $Z$ on the charge distribution of $\Psi_X$.

 For X$^*$, the variational function is:\cite{ClimenteJPCc}
\begin{eqnarray}
	\Psi_{X^*} 
	&=&   
	\Phi_e(\mathbf{r}_{e1},\mathbf{r}_{e2}) \,
	\Phi_h(\mathbf{r}_{h}) \, 
	J(r_1,r_2,r_{12})\, 
	\label{eq:PsiX*}
\end{eqnarray}
\noindent with $\Phi_h(\mathbf{r}_h)$ and $\Phi_e(\mathbf{r}_{e1},\mathbf{r}_{e2})$ the analytical non-interacting-particle wave function, 
and $J(r_1,r_2,r_{12})$ the following Jastrow factor:
\begin{equation}
\label{eq:J}
	J(r_1,r_2,r_{12}) = e^{-Z s/2} \, \cosh{(Z Q t /2)} \, 
	e^{\frac{Z\, b \, r_{12}}{(1+Z\, a \, r_{12})}}.
\end{equation}
\noindent In this expression, $r_{1(2)}=|\mathbf{r}_{e1(2)}-\mathbf{r}_h|$,  $r_{12}=|\mathbf{r}_{e1}-\mathbf{r}_{e2}|$, $s=r_1+r_2$ and $t=r_1-r_2$.
The first term of $J$ is a Slater correlation factor for electrons and holes, similar to the X case. 
$Z$ is then a variational parameter, related to the electron-hole interaction strength.
Its influence is depicted in Figure \ref{fig1}(b).
Larger $Z$ values imply shorter electron-hole distances.
The second term is to enable asymmetric electron-hole interaction for the two electrons.
As can be seen in the following relation:
\begin{equation}
e^{-Z s/2} \, \cosh{(Z Q t /2)} = (e^{-Z_1 r_1 - Z_2 r_2} + e^{-Z_1 r_2 - Z_2 r_1})/2,
\end{equation}
\noindent 
where $Z_1=Z(1 + Q)/2$ and $Z_2 = Z (1 - Q)/2$,
a non-zero $Q$ value yields different interaction strengths.
 A sketch of the influence of $Q$ in the trion charge distribution is shown in Fig.~\ref{fig1}(c).
 When increasing from $Q=0$ to $Q=1$, the system evolves from perfectly bound trion towards a dissociated exciton-electron complex. 
This flexibility of the wave function has been found to play an important role for trions in layered halide perovskites,
where one electron-hole pair stays within the polaronic radius, while the extra carrier orbits farther away.\cite{ClimenteJPCc}
$Q$ is then our second variational parameter for X$^*$.
The last term in $J$ is a Pad\'e Jastrow factor, which has the property of giving the desired limits with $r_{12}$. 
At short ranges of interaction, $r_{12}\rightarrow 0$, the term becomes $e^{b\,r_{12}}$, 
which provides a cusp to compensate for the divergence in electron-electron Coulomb repulsion ($b>0$).
At the same time, the probability to find distant holes ($r_{12} \rightarrow \infty$)
is more likely than that of proximal holes ($r_{12} \rightarrow 0$) by a factor $(e^{b/a})^2$. 
 We take $b$ and $a$ as variational parameters. \\

 For BX, our starting variational wave function is formally the same we have successfully employed
 in the past to study nanoplatelets,\cite{ClimenteJPCL,MaciasNS}
\begin{eqnarray} 
        \Psi_{BX} 
	= \Phi_{e}(\mathbf{r}_{e_1})\Phi_{e}(\mathbf{r}_{e_2})\Phi_{h}(\mathbf{r}_{h_a}) \Phi_{h}(\mathbf{r}_{h_b})  \nonumber  \\
        \; F(r_{1a},r_{1b},r_{2a},r_{2b},r_{12},r_{ab}).
	\label{twf}
\end{eqnarray}
Here, 
 $F$ is the correlation factor, described by:
\begin{eqnarray}
F
=e^{-Z\frac{(s_1+s_2)}{2}} \cosh\Big(ZQ\frac{t_1-t_2}{2}\Big)  \nonumber  \\
\times e^{Z \frac{\beta r_{12}}{1+Z\alpha r_{12}}} e^{Z \frac{\beta r_{ab}}{1+Z\alpha r_{ab}}},
\label{fcor}
\end{eqnarray}
\noindent where $s_1=r_{1a}+r_{1b}$, $s_2=r_{2a}+r_{2b}$, $t_1=r_{1a}-r_{1b}$ and $t_2=r_{2a}-r_{2b}$ with $r_{12}, r_{ab}, r_{1a}, r_{1b}, r_{2a}, r_{2b}$ the interparticle distances. $Z$ (electron-hole interaction strength), $Q$ (degree of X-X dissociation), $\beta$ and $\alpha$ (repulsion correlation terms) are the variational parameters to be optimized.
 For the materials and geometries we consider, $Z$ and (to a lesser extent) $Q$ parameters are found to be major contributions on the energy. 
 Their qualititive influence on $\Psi_{BX}$ is sketched in Figs.~\ref{fig1}(d)
 and \ref{fig1}(e). 
 As shown in Fig.~\ref{fig1}(d), increasing $Z$ implies reducing electron-hole distances, and hence the size of the BX.  
 As shown in Fig.~\ref{fig1}(e), increasing $Q$ (from $Q=0$ to $Q=1$), dissociates the two X forming the BX.
 \\

In some calculations, to explore if $\Psi_{BX}$ misses electronic correlations, we introduce additional degrees of freedom. 
The first one takes into account that Eq.~(\ref{fcor}) is restricted to a symmetric excitonic dissociation path. 
That is to say, as $Q$ grows from $Q=0$ to $Q=1$, the BX dissociates into two identical Xs, 
both with the same electron-hole attraction ($Z$). 
We can relax this restriction with an extra parameter $\delta Z$ in one of the excitons, 
so that, while preserving the exchange of identical particles,  
we set $z_1 = Z$, $z_2 = Z + \delta Z$. 
%
 The complete correlation factor becomes:
%
\begin{eqnarray}
	F = \frac{1}{2} \left[ e^{-\frac{z_1 s_1+ z_2 s_2}{2}} \cosh{(\frac{z_1 Q \, t_1- z_2 Q \,t_2}{2})} + \nonumber \right. \\
	\left. e^{-\frac{z_2 s_1+ z_1 s_2}{2}} \cosh{(\frac{z_2 Q \, t_1- z_1 Q \,t_2}{2}])} \right]  \nonumber \\ 
	e^{-Z \frac{\beta r_{12}}{1+Z \alpha r_{12}}} \; e^{-Z \frac{\beta r_{ab}}{1+Z \alpha r_{ab}}}.
\label{eq5}
\end{eqnarray}
\noindent This function allows one to break the symmetry in the attractions while keeping unchanged the parameters for repulsion. 
	We shall refer to $\delta Z$ as the microscopic asymmetry factor.
	Its influence on $\Psi_{BX}$ is sketched in Fig.~\ref{fig1}(f).
	\\

The second degree of freedom we add will be referred to as the macroscopic asymmetry factor, $\lambda$.
The goal in this case is to reduce the overlap of identical particles while keeping it unchanged for the different ones. 
To this end, we do not modify the correlation factor, but rather the envelope functions, $\Phi_{e}$  and $\Phi_{h}$.
For electrons, the replacement is:
\begin{eqnarray}
\label{eq6}
	&&	\cos{(k x_1)} \cos{(k x_2)} \to  \nonumber  \\
	&&	\left[ \cos{(k x_1)}+\lambda \sin{(2 k x_1)} \right] \left[ \cos{(k x_2)}-\lambda \sin{(2 k x_2)} \right], \nonumber
\end{eqnarray}
\noindent and for holes:
\begin{eqnarray}
	&& \cos{(k x_a)} \cos{(k x_b)}  \to \nonumber \\
	&& \left[ \cos{(k x_a)}+\lambda \sin{(2 k x_a)} \right] \left[\cos{(k x_b)}-\lambda \sin{(2 k x_b)} \right], \nonumber
\end{eqnarray}
\noindent with $k =\pi / L$. We proceed in a similar way for $y$ and $z$ coordinates.
The qualitative effect of $\lambda$ is pictorially sketched on Fig.~\ref{fig1}(g).
The overlap between $e_1$ and $h_a$ ($e_2$ and $h_b$) remains unity, but that
between identical particles ($e_1$ and $e_2$, $h_a$ and $h_b$) becomes
$(\frac{1-\lambda^2}{1+\lambda^2})^3$.  
%
 %

\section{Results}

We describe CsPbBr$_3$ NCs using the material parameters 
extracted in Ref.~\onlinecite{BaranowskiACSener}. 
Thus, bare electron and hole masses are 
$m_e^b=m_h^b=0.234\,m_0$, with $m_0$ the free electron mass.
 The Fr\"olich coupling constant is $\alpha=2.1$,
 which gives polaronic masses
$m_e=m_h=0.316\,m_0$.
 The LO phonon energy is $E_{LO}=18$ meV.
 From Eq.~(\ref{eq:l}), the previous parameters yield polaron radii $l_e = l_h = 3$ nm.
Relative static and dynamic dielectric constants inside the NC ($\varepsilon_{NC}$) 
are $\varepsilon_{s} = 16$ and $\varepsilon_{\infty} = 4.5$.
  The dielectric constant of the surrounding matrix (polystyrene)
 is set to $\varepsilon_{out} = 2.56$.\cite{ZhuAM} 
 The band gap is taken from the X emission in the limit of bulk, $E_{gap}=2.361$ eV.\cite{ZhuNAT}
 In order to retrieve the bulk X binding energy, $\Delta_X^{bulk}=33$ meV,\cite{ZhuNAT,YangACSener}
  the Bajaj optimization parameter is set here to $\gamma=3/13$. 
 This value grants an optimal description of the polaronic exciton in bulk CsPbBr$_3$, 
 upon which the subsequent influence of interactions with other carriers and with  confinement will be studied. 
 Using $\gamma=3/5$, which is the (material-averaged) value proposed in Ref.~\onlinecite{BajajSSC}
 for a number of different ionic crystals, 
 underestimates the bulk exciton binding energy ($\Delta_X^{bulk} (\gamma=3/5)=19$ meV) 
 and does not change the conclusions we will derive. 
 We have checked that Haken and Pollmann-B\"uttner potentials also fail to reproduce $\Delta_X^{bulk}$.
 %
  \\

\begin{figure}[h]
\includegraphics[width=8.0cm]{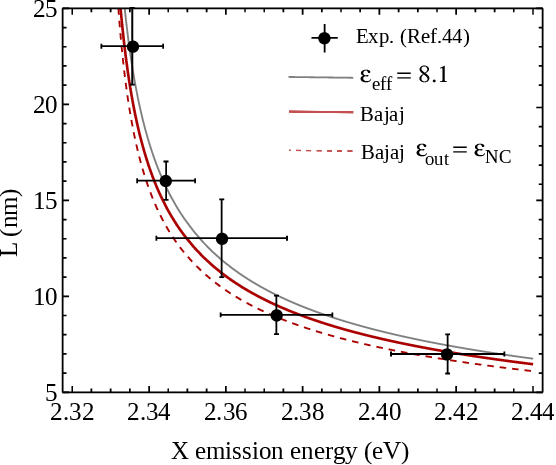}
\caption{Emission energy of cuboidal CsPbBr$_3$ NCs with different lateral sizes. 
	Calculations under different approximations are compared with experimental values (see text).
	They all provide reasonable agreement}
\label{fig2}
\end{figure}

 To test the validity of our model, we start by comparing the calculated X energies
 with the experimental band edge emission of CsPbBr$_3$ NCs reported in
 Ref.~\onlinecite{ZhuNAT}.
 Figure \ref{fig2} shows the experimental data (black dots) alongside our simulations, 
 for NCs of different sizes. 
 A solid red line is used to represent the results of the full model, 
 including distance-dependent dielectric screening.
 For comparison, we also show: (i) 
 results neglecting dielectric confinement ($\varepsilon_{out}=\varepsilon_{NC}$), 
 and (ii) results neglecting the different screening at short and long distances 
 (i.e. replacing Eq.~\ref{eq02} by a simple Coulomb interaction term). 
 The latter approximation has been used in recent theoretical studies, 
 which took an effective dielectric constant in between the 
 static and dynamic limit.\cite{ChoACS,NguyenPRB,ZhuNAT,ZhuAM} 
 The effective constant is chosen so as to fit the bulk binding energy of CsPbBr$_3$. 
 With our masses, this yields $\varepsilon_{eff}=8.1$.
 It is inferred from Figure \ref{fig2} that, on this energy scale, 
 all levels of approximation provide reasonable agreement with experimental data. 
 This is in contrast to nanoplatelets and layered perovskites, 
 where the strong quantum confinement leads to conspicuous dielectric mismatch and 
 short-distance (Yukawa) interactions.\cite{MovillaNSa}
 CsPbBr$_3$ NCs with the sizes we study  ($L \geq 5$ nm) are in the mid- to weak confinement regime.\cite{ButkusCM}
 The X ground state is then farther from the polystyrene region, which implies weaker dielectric confinement,
 and has larger size, which implies less pronounced short-distance interactions.

\begin{figure}[h]
\includegraphics[width=8.0cm]{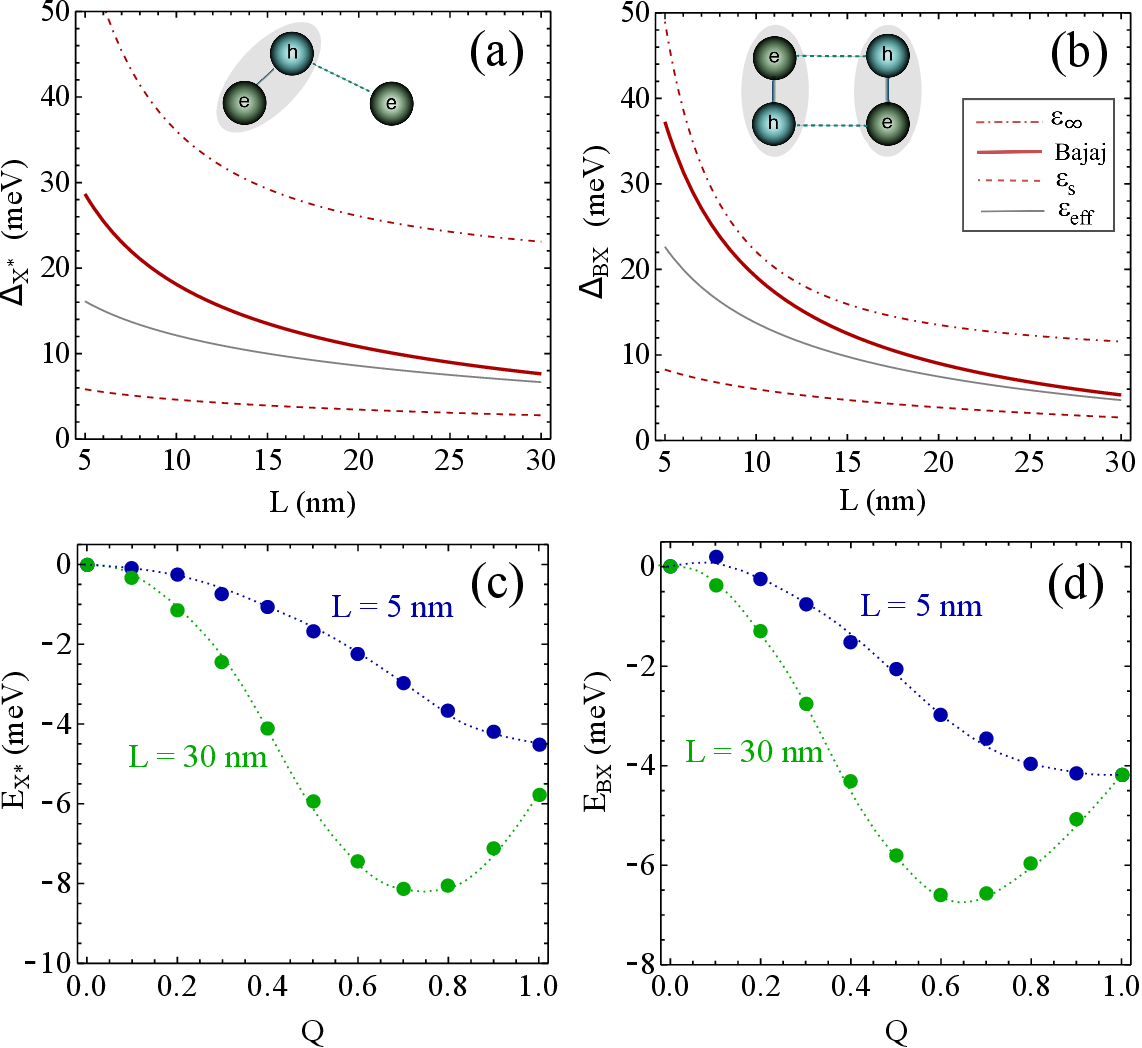}
\caption{
(a) and (b): size dependence of X$^*$ and BX binding energy. 
Red, thick lines: Bajaj potential. 
Dashed lines: zero polaron radius limit. 
Dot-dashed lines: infinite polaron radius limit. 
Gray lines: Coulomb potential with effective screening ($\varepsilon_{eff}=8.1$). 
Insets: Schematics of the process of X$^*$ and BX formation. 
(c) and (d): X$^*$ and BX energy dependence on the (dimensionless) variational parameter $Q$, 
showing the extra stabilization of non-symmetric arrangements of the charge distribution. 
The origin of energies in each panel corresponds to that of the most symmetric structures ($Q = 0$). 
Lines are guides to the eye. 
}
\label{fig3}
\end{figure}

We next study the electronic structure and charge distribution of X$^*$ and BX, as derived from our model.
The following relevant energy differences are defined:
\begin{equation}
	\Delta_{X^*} = (E_X + E_e) - E_{X^*}, 
\end{equation}
\noindent and 
\begin{equation}
	\Delta_{BX} = 2\,E_X - E_{BX},
\end{equation}
\noindent where $E_i$ is the ground state energy of the species $i$.
These expressions closely correspond to the spectral shift between X$^*$ and X (BX and X) 
resonances in the emission spectra of NCs.
 They are often identified with binding energies, which is strictly true in the bulk limit.\cite{ClimenteJPCc}
Positive values of $\Delta_{X^*}$ and $\Delta_{BX}$ stand for bound (redshifted) complexes.
Figure \ref{fig3}(a) and (b) show the calculated values 
in NCs of different lateral side $L$.
Red solid line represents the full calculation, discriminating 
short and long-distance interactions. 
For comparison, we also show the binding energies obtained with a Coulomb potential instead,
using either $\varepsilon_s$ (red dashed line) or $\varepsilon_\infty$ (red dashed-dotted line) screening.
In all approximations, $\Delta_X^*$ and $\Delta_{BX}$ increase with quantum confinement,
as expected from the stronger carrier-carrier interactions in reduced space. 
 The Bajaj potential gives intermediate values as compared to the static and dynamic screening limits. 
For large NCs, the energies are closer to the static limit,
but they evolve towards the dynamic one as the NC becomes smaller.
The approach to the dynamic limit is remarkable for BX, which agrees with previous, more elaborated 
estimates of $\Delta_{BX}$ in the smaller NC range.\cite{ParkPRM} 
Likewise, using a Coulomb potential with effective screening $\varepsilon_{eff}$ (gray line in the figure),
as in Refs.~\cite{ChoACS,NguyenPRB,ZhuNAT,ZhuAM}, is a good approximation for large NCs, but it underestimates $\Delta_{X^*}$ and $\Delta_{BX}$ by
a significant amount when confinement increases (a factor of $0.3-0.4$ for $L=5$ nm) . 
It follows that the distance-dependent screening of polaronic interactions plays a significant role
in determining binding energies of CsPbBr$_3$ NCs. 

In our calculations, the ground state of X$^*$ and BX is found to present significant charge distribution asymmetries.
This can be seen in Figure \ref{fig3}(c) and (d), which illustrate the influence of the variational parameter $Q$
for small ($L=5$ nm) and large ($L=30$ nm) NCs.
For both species and all NC dimensions, the energy minima are found at non-zero values of $Q$.
As discussed in Figure \ref{fig1}(c,e), this corresponds to partially dissociated excitonic complexes.
Trions are then formed by an exciton plus a more distant, orbiting electron.
This can be seen as a sizable ionic, as opposed to covalent, bond character.
In turn, BXs are formed by two $Xs$, with X-X interactions being weaker than intra-X ones.
This behavior is reminiscent of that found in quasi-2D lead halide perovskites.\cite{ClimenteJPCc,ClimenteJPCL}
The absolute energy of the asymmetry corrections is smaller in NCs ($<10$ meV), 
and yet it provides a non-neglibible contribution to $\Delta_{X^*}$ and $\Delta_{BX}$.\\

\begin{figure}[h]
\includegraphics[width=8.0cm]{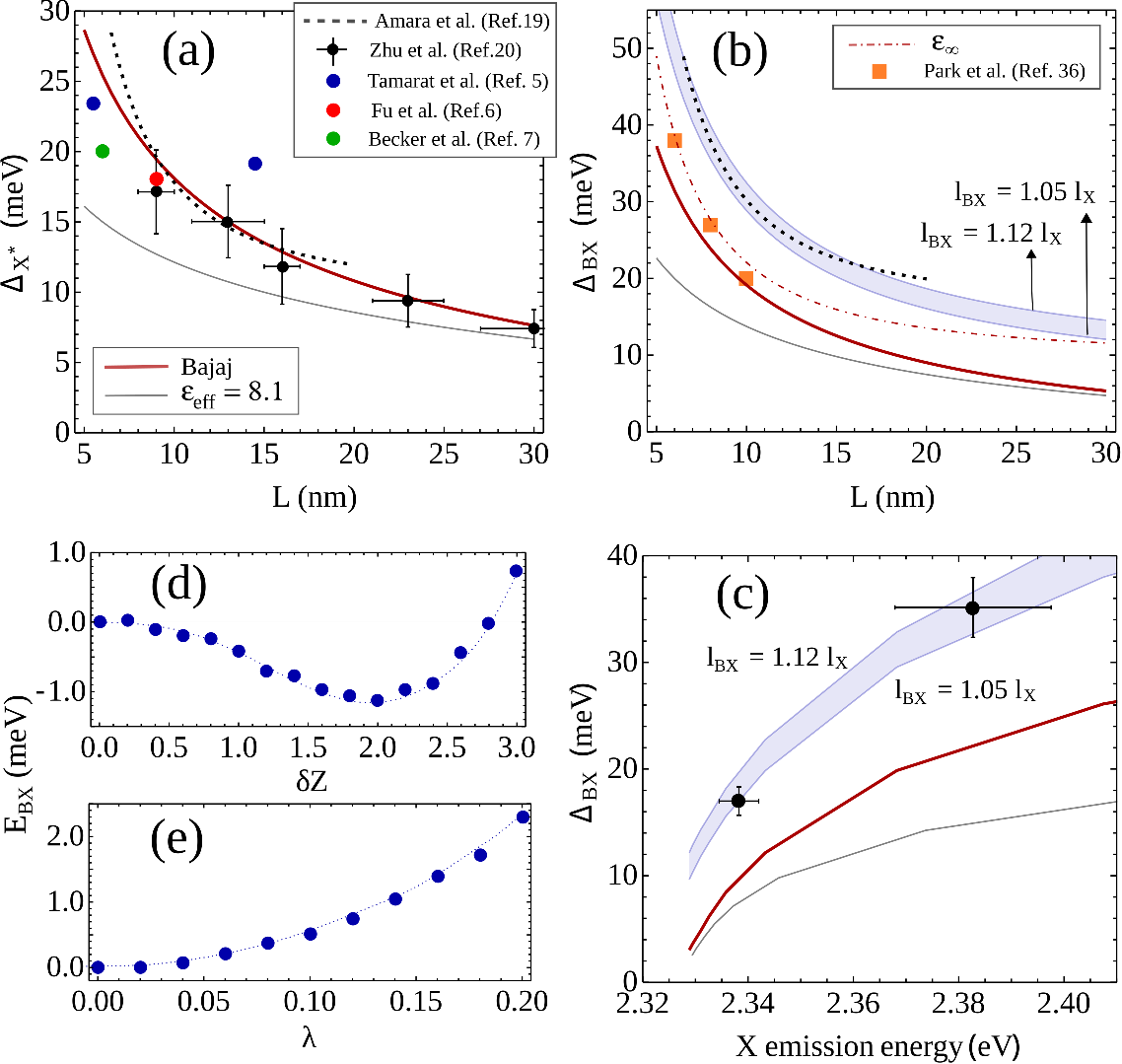}
\caption{
	(a)	Size dependence of the calculated X$^*$ binding energy (solid lines) compared to experimental data (dots, dotted line). The modified Bajaj model (red line) reproduces well the experiments. The effective screening model (gray line) underestimates the value.  (b,c) Same for the BX. Experimental BX shifts from Amara et al.~\cite{AmaraNL} and Zhu et al.\cite{ZhuAM} are depicted by the dashed line and the black dots, respectively. Orange squares correspond to theoretical estimates from Park et al.~\cite{ParkPRM} The figure also includes the limit of dynamic screening (dot-dashed line) as well as results assuming a 5 to 12\% increase of the polaron radii in the BX regime (blue shaded area). (d) and (e) BX energy dependence on the variational parameters $\delta Z$ and $\lambda$, for a NC with $L=30$ nm. The other four parameters are fixed at their optimized values for $\delta Z=\lambda=0$.
}
\label{fig4}
\end{figure}

Having analyzed the influence of polaronic (Haken-like) terms on the size-dependence 
of $\Delta_{X^*}$ and $\Delta_{BX}$, we next compare the theoretical estimates
with experimentally measured values.
Figure \ref{fig4}(a) shows the comparison for X$^*$. 
The calculations using a distance-dependent screening (red solid line) are in very 
good agreement with data points from different experiments\cite{FuNL,ZhuAM,TamaratNC,BeckerNAT},
as well as with the trend inferred from the systematic experiments of Amara and co-workers.\cite{AmaraNL}
Using an effective dielectric constant (gray line), however, underestimates the binding energies in small NCs.
This is indicative that, in lead halide NCs, non-hydrogenic polaronic interactions can make a significant difference, 
as observed earlier in bulk\cite{BaranowskiAEM,MenendezPSS,BaranowskiACSener} and layered structures.\cite{MovillaNSa,ClimenteJPCL,ClimenteJPCc}

A completely different conclusion is drawn in the case of BXs.
As shown in Figures \ref{fig4}(b) and \ref{fig4}(c), $\Delta_{BX}$ calculated with the Haken-like model
(red solid line) falls short as compared to experimental values measured in Refs.~\onlinecite{AmaraNL,ZhuAM}.
This is inspite of the binding energies being larger than those obtained with 
an effective dielectric constant (gray line).
The discrepancy is persistent for all $L$ values.
It remains even when non-parabolic (heavier) masses\cite{SercelJCP} are used for electrons and holes (not shown), 
or when larger polaron radii are assumed 
(see e.g. the dynamic limit, dot-dashed line in Fig.~\ref{fig4}(b)). 
%
To verify that the origin is not related to insufficient correlation energy in our simulations, 
we modify our variational wave function, Eq.~(\ref{twf}), by introducing two additional 
variational parameters $\delta Z$ and $\lambda$. 
As explained in Section \ref{s:theo}, $\delta Z$ introduces asymmetry through the correlation factor
and $\lambda$ through the envelope functions.
Figure \ref{fig4}(d) and (e) show that neither of these factors are able to lower $E_{BX}$ significantly. 
As a matter of fact, $\lambda \neq 0$ is only making the BX less stable, because the
increase in kinetic energy exceeds the benefit of reduced electron-electron and hole-hole repulsions.

It is worth noting that our $\Delta_{BX}$ estimates are close to those obtained using alternative 
(atomistic) models of polaronic effects, dielectric confinement and electronic correlations 
(orange squares in Fig.~\ref{fig4}(b)),\cite{ParkPRM} which also underestimate the energies as compared to experiments.

Having ruled out polaronic interactions and electronic correlations as the origin of the large 
$\Delta_{BX}$ in NCs, we can speculate on the origin of the large experimental value of $\Delta_{BX}$.
 For example, in Haken-like models, originally developed for bulk, 
the polaron radius depends on $E_{LO}$ --recall Eq.~(\ref{eq:l}).
Unlike bulk, however, NCs present low-energy phonon modes which
may be related to LO phonons as well.\cite{AmaraNL,ZhuAOM}
If the averaged $E_{LO}$ value turns out to be smaller for BX than for X,
 $l_{BX} > l_X$  may be expected. As shown in Fig.~\ref{fig4}(b) --blue shaded area--, 
 an increase of $l_{BX}$ in 5-12\% would suffice to match experimental $\Delta_{BX}$ values.
This possibility is in line with --but different from-- previous proposals, 
such as the effective dielectric constant felt by BX being lower than that of X.\cite{ZhuAM}
 A different $E_{LO}$ for X and BX may be expected because they have different sizes and charge distribution.
 This implies different degree of interaction with the nanocrystal surface,
 which may play a role in the formation of low-energy optical phonon modes,
 and with the phonon modes themselves.
 Experimentally, this should be seen as the replica having different enegy shifts or intensity in each excitonic complex.
%
%
 Either way, we need to assume that X and BX polarize the perovskite lattice differently.
 This seems to agree with the conclusions of recent time-resolved, 
 optical-pump--electron-diffraction-probe experiments
 pointing out that the carrier-phonon coupling strength 
in lead halide perovskite lattices scales quadratically 
with the exciton number, so that it is stronger for BX
than for X or X$^*$.\cite{YazdaniNP}
Further investigations will be needed to establish the actual mechanism for the large $\Delta_{BX}$.

\section{Conclusions}

We have investigated the influence of non-hydrogenic polaronic 
interactions in X, X$^*$ and BX confined in CsPbBr$_3$ NCs, 
by means of a Haken-like (Bajaj) potential in an effective mass - variational Quantum Monte Carlo model.
The X$^*$ ground state is formed by an exciton plus a more distant, 
orbiting electron. The BX ground state, in turn, is formed by two 
X with inter-X interaction being weaker than intra-X one.
These charge distributions are reminiscent of those in layered halide perovskites.\cite{ClimenteJPCL,ClimenteJPCc}

It has been shown that distance-dependent dielectric screening 
(a signature of polaronic interactions in soft lattice materials)
enhances $\Delta_{X^*}$ and $\Delta_{BX}$ in CsPbBr$_3$ NCs.
Nonetheless, for all the NC sizes we study, $\Delta_{BX}$ 
is significantly smaller than that inferred from
spectral shifts in single-particle experiments at cryogenic temperatures.\cite{AmaraNL,ZhuAM,FuNL,BeckerNAT,TamaratNC,ChoACS}
That is, neither polaronic interactions, nor dielectric confinement,
nor electronic correlations (all of which are well captured in our model) 
suffice to explain the large magnitude of $\Delta_{BX}$.
This result suggests that BXs and Xs may polarize the lattice differently,
 leading to different dielectric screening and/or to different 
 renormalization of the band gap upon formation of polarons.\\

\begin{acknowledgments}
We are grateful to G. Rain\'o for discussions.
We acknowledge support from Grant No. PID2021-128659NB-I00, funded by Ministerio de Ciencia e Innovaci\'{o}n (MCIN/AEI/10.13039/501100011033 and ERDF ``A way of making Europe'').
\end{acknowledgments}

\bibliography{pv_nc}

\end{document}